\documentclass[twocolumn,showpacs,showkeys,preprintnumbers,amsmath,amssymb]{revtex4-2}
\usepackage[latin9]{inputenc}
\usepackage{amsmath}
\usepackage{amssymb}
\usepackage{graphicx}
\usepackage{ulem}
\usepackage{xcolor}

\begin{document}




\newcommand*\mycommand[1]{\texttt{\emph{#1}}}

\author{G. Barbero$^{1,2}$, L. R. Evangelista$^{1,3}$, E. K. Lenzi$^{4}$}
\affiliation{$^1$Dipartimento di Scienza Applicata del Politecnico di Torino,\\
Corso Duca degli Abruzzi 24, 10129 Torino, Italy.\\
$^2$National Research Nuclear University MEPhI (Moscow Engineering Physics Institute), Kashirskoye shosse 31, 115409 Moscow, Russian Federation.\\
$^3$ Departamento de F{\'i}sica, Universidade Estadual de Maring\'a, Avenida Colombo, 5790, 87020-900 Maring\'a, Paran\'a, Brazil. \\
$^4$  Departamento de F{\'i}sica,  Universidade Estadual de Ponta Grossa, Avenida Carlos Cavalcanti, 4748, 87030-900, Ponta Grossa, Paran\'a, Brazil.}
\email{E-Mails:\\giovanni.barbero@polito.it (G.B.)\\ lre@dfi.uem.br (L.R.E.)\\ eklenzi@uepg.br (E.K.L.)}

\title[CV]
  {Time-Fractional Approach to the Electrochemical Impedance: The Displacement Current}




\begin{abstract}
We establish, in general terms, the conditions to be satisfied by a time-fractional approach formulation of the Poisson-Nernst-Planck model in order to guarantee that the total current across the sample be solenoidal, as required by the Maxwell equation.  Only in this case the electric impedance of a cell can be determined as the ratio between the applied difference of potential and the current across the cell. We show that in the case of anomalous diffusion, the model predicts for the electric impedance of the cell a constant phase element behaviour in the low frequency region. In the parametric curve of the reactance versus the resistance, the slope coincides with the order of the fractional time derivative.
\end{abstract}
\maketitle

The Poisson-Nernst-Planck (PNP) diffusional model is formulated to consider a neutral species that can dissociate into positive and negative charges of arbitrary mobilities~\cite{Ross1953,Trukhan1963}. In the presence of the external electric field $\mathbf{E}({\textbf{r}},t)$, these charges move, giving rise to currents of neutral as well as of  positive and negative charges. If we denote the bulk density of these particles, respectively with $n_n ({\textbf{r}}, t)$, $n_p({\textbf{r}}, t)$, and $n_m({\textbf{r}}, t)$ and, likewise, the current density as ${\textbf{j}}_n ({\textbf{r}}, t)$, ${\textbf{j}}_p({\textbf{r}}, t)$, and ${\textbf{j}}_m({\textbf{r}}, t)$, the continuity equations will be written as
\begin{eqnarray}
\label{Continuity-normal}
\frac{\partial}{\partial t} n_p({\textbf{r}}, t) &=& - \nabla \cdot {\textbf{j}}_p({\textbf{r}}, t) + S({\textbf{r}}, t), \nonumber \\
\frac{\partial }{\partial t}n_m({\textbf{r}}, t) &=& - \nabla \cdot {\textbf{j}}_m({\textbf{r}}, t) + S({\textbf{r}}, t),\nonumber \\
\frac{\partial }{\partial t}n_n({\textbf{r}}, t) &=& - \nabla \cdot {\textbf{j}}_n({\textbf{r}}, t) - S({\textbf{r}}, t),
\end{eqnarray}
where $S({\textbf{r}}, t)$ accounts for a source term related to the dissociation of neutral particles and recombination of the ions. In the case of a full dissociation the neutral particles desappear, as well as the source term.    To obtain the set of the fundamental equations of the PNP model, we have to consider also the Poisson equation, written as
\begin{equation}
\label{Eq-Poisson}
\nabla \cdot \mathbf{E}({\textbf{r}},t) = \frac{q}{\varepsilon} \left[n_p({\textbf{r}}, t)- n_m({\textbf{r}}, t) \right],
\end{equation}
where $\varepsilon$ is the dielectric permittivity of the insulating medium free of ions.  This equation connects the bulk density of ions of positive and negative charges, of absolute value $q$,  to the actual profile of the electric field across the sample.

The total electric current is formed by the conduction and the displacement currents:
\begin{equation}
\label{Current-Total}
{\textbf{j}}({\textbf{r}}, t) = q \left[{\textbf{j}}_p({\textbf{r}}, t) - {\textbf{j}}_m({\textbf{r}}, t)\right] + \varepsilon \frac{\partial \mathbf{E}({\textbf{r}},t) }{\partial t}.
\end{equation}
If we now combine the first two of Eqs.~(\ref{Continuity-normal}) with
Eq.~(\ref{Current-Total}), we conclude that
\begin{equation}
\nabla \cdot {\textbf{j}} = -q \left\{\frac{\partial}{\partial t} \left[n_p({\textbf{r}}, t)-n_m({\textbf{r}}, t)\right]\right\} + \varepsilon \nabla \cdot \frac{\partial \mathbf{E}({\textbf{r}},t) }{\partial t}=0,
\end{equation}
when the equation of Poisson, Eq.~(\ref{Eq-Poisson}),  is used. Indeed, the total current has to be solenoidal, i.e,
\begin{equation}
\label{Jzero}
\nabla \cdot {\textbf{j}}({\textbf{r}}, t) = \nabla \cdot \left\{q \left[{\textbf{j}}_p({\textbf{r}}, t) - {\textbf{j}}_m({\textbf{r}}, t) \right] + \varepsilon  \frac{\partial \mathbf{E}({\textbf{r}},t) }{\partial t}\right\}=0.
\end{equation}
In this case, for a one-dimensional problem, the current density $\bf j$ is position independent.
Only in this framework, the concept of electrical impedance, defined as the ratio between the difference of potential applied to the system and the total electric current flowing across it, is meaningful \cite{ioannis2017}.
To proceed with the application of the PNP model to deal with experimental data obtained in electrolytic cells, the set of Eqs.~(\ref{Continuity-normal}) and~(\ref{Eq-Poisson}) has to be solved for well-defined boundary conditions. Anyway, the total  current flowing though the cell is subjected to the general condition established in Eq.~(\ref{Jzero}), i.e., the total current has to be solenoidal.

Perhaps the simplest extension of the problem using derivatives operators of arbitrary order, commonly called fractional derivatives,  should read as the one we have proposed some years ago~\cite{JPCB113}:
\begin{eqnarray}
\label{Continuity-frac}
\tau^{\gamma-1}\frac{\partial^{\gamma}n_p({\textbf{r}}, t)}{\partial t^{\gamma}}  &=& - \nabla \cdot {\textbf{j}}_p({\textbf{r}}, t) + S({\textbf{r}}, t), \nonumber \\
\tau^{\gamma-1}\frac{\partial ^{\gamma}n_m({\textbf{r}}, t) }{\partial t^{\gamma}} &=& - \nabla \cdot {\textbf{j}}_m({\textbf{r}}, t) + S({\textbf{r}}, t),\nonumber \\
\tau^{\gamma-1}\frac{\partial ^{\gamma} n_n({\textbf{r}}, t) }{\partial t^{\gamma}} &=& - \nabla \cdot {\textbf{j}}_n({\textbf{r}}, t) - S({\textbf{r}}, t),
\end{eqnarray}
where $\tau$ is an intrinsic time of the problem. In Eqs.~(\ref{Continuity-frac})  we have replaced the time derivative
present in Eqs.~(\ref{Continuity-normal}) with the Riemann-Liouville
fractional derivative, by promoting the change:
\begin{equation}
\frac{\partial n_{p,m} ({\textbf{r}},t)}{\partial t} \longrightarrow \frac{\partial^{\gamma} n_{p,m} ({\textbf{r}},t)}{\partial t^{\gamma}} \longrightarrow {_{t_{0}}\textrm{D}}_{t}^{\gamma} n_{p,m}
({\textbf{r}},t),
\end{equation}
and used the definition of the Riemann-Liouville operator as
\begin{eqnarray}
\!\!\!\!\!\!\;_{t_{0}}{{\textrm{D}}}_{t}^{\gamma} n_{p,m}
({\textbf{r}},t)=\frac{1}{\Gamma\left(m-\gamma\right)}\frac{d^{m}}{dt^{m}}\int_{t_{0}}^{t}d\overline{t}\frac{
n_{p,m}({\textbf{r}},\overline{t}\;)}{\left(t-\overline{t}\right)^{\gamma+1-m}},
\end{eqnarray}
where $m-1<\gamma<m$, with $m$ an integer,  and $t_0$ is related to the conditions
initially imposed to the system. For an analysis of the kind we are proposing here, focused on the impedance spectroscopy,  without loss of generality we can consider
$t_{0}=-\infty$ to study the response of the system to the
periodic applied potential,  as required by the impedance spectroscopy technique. If we now combine the new set of fundamental equations of the extended model, which are Eqs.~(\ref{Continuity-frac}) and (\ref{Eq-Poisson}), following the procedure used before, it is mandatory to conclude that
$ \nabla \cdot {\textbf{j}}({\textbf{r}},t) = 0$, i.e.,
\begin{equation}
\label{Jzero-frac}
\nabla \cdot \left\{q \left[{\textbf{j}}_p({\textbf{r}},t) - {\textbf{j}}_m({\textbf{r}},t)\right] + \varepsilon \tau^{\gamma-1}  \frac{\partial^{\gamma} \mathbf{E}({\textbf{r}},t) }{\partial t^{\gamma}}\right\}=0.
\end{equation}
This means that a physically sound extension of the PNP model using fractional derivatives requires an extended expression in which the displacement current is also defined in terms of a derivative of arbitrary order. This fact has profound implications and requires discussing also possible extensions of the Maxwell equations to the field of the differential operators of arbitrary order. Before analyzing this important aspect of the problem, to keep the approach as general as possible, let us consider a further step in the generalization of the PNP model using the time-fractional derivative.

One possible way to generalize the previous equations is to use the so-called time-fractional derivative operator of distributed order, defined as~\cite{Chechkin,Mainardi}:
\begin{equation}
\int_{0}^{1}d\gamma p(\gamma)
\!\;_{-\infty}{\rm D}_{t}^{\gamma}\left(\cdots
\right),
\end{equation}
where $p(\gamma)$ is a distribution of
$\gamma$, with
$\int_{0}^{1}d\gamma
p(\gamma)=1$,
instead of using only a single fractional operator $_{-\infty}{\textrm{D}}_{t}^{\gamma}\left(\cdots
\right)$. After implementing this generalization, the set of fundamental equations of the extended PNP model will be now formed by the following continuity equations:
\begin{eqnarray}
\label{Continuity-fractional1}
\!\!\!\!\!\!\!\!\int_{0}^{1}\!\!d\gamma p(\gamma)\tau^{\gamma-1}\frac{\partial^{\gamma} }{\partial t^{\gamma}}n_p({\textbf{r}}, t) &=& - \nabla \cdot {\textbf{j}}_p({\textbf{r}}, t) + S({\textbf{r}}, t), \nonumber \\
\!\!\!\!\!\!\!\!\int_{0}^{1}\!\!d\gamma p(\gamma)\tau^{\gamma-1}\frac{\partial^{\gamma} }{\partial t^{\gamma}}n_m({\textbf{r}}, t) &=& - \nabla \cdot {\textbf{j}}_m({\textbf{r}}, t) + S({\textbf{r}}, t), \nonumber \\
\!\!\!\!\!\!\!\!\int_{0}^{1}\!\!d\gamma p(\gamma)\tau^{\gamma-1}\frac{\partial^{\gamma} }{\partial t^{\gamma}}n_n({\textbf{r}}, t) &=& - \nabla \cdot {\textbf{j}}_n({\textbf{r}}, t) - S({\textbf{r}}, t),
\end{eqnarray}
and the equation of Poisson, Eq.~(\ref{Eq-Poisson}). Combining again, as before, these equations, it is possible to show that
\begin{eqnarray}
\int_{0}^{1}d\gamma p(\gamma)\!\!\!\!\!\!\!&&\tau^{\gamma}\frac{\partial^{\gamma} }{\partial t^{\gamma}}\left[n_p({\textbf{r}}, t)-n_m({\textbf{r}}, t) \right] \nonumber \\ &+& \nabla \cdot \left[{\textbf{j}}_p({\textbf{r}}, t)-{\textbf{j}}_{m}({\textbf{r}},t)\right]=0,\label{total_Current_Fractional}
\end{eqnarray}
and, consequently, that
\begin{eqnarray}
\label{Jzero-order}
 \nabla \! \cdot \! \left[q\left[{\textbf{j}}_p({\textbf{r}}, t) - {\textbf{j}}_m({\textbf{r}}, t) \right]\!+\!
 \frac{\varepsilon}{\tau} \int_{0}^{1}\!\!\!\!d\gamma p(\gamma)\tau^{\gamma}\frac{\partial^{\gamma} }{\partial t^{\gamma}}\mathbf{E}({\textbf{r}},t)\right]=0.\nonumber\\
\end{eqnarray}
Equation~(\ref{Jzero-order}) is now more general than Eq.~(\ref{Jzero-frac}) and implies that the total electric current
\begin{eqnarray}
{\textbf{j}}({\textbf{r}},t)&=&  q \left[{\textbf{j}}_p({\textbf{r}}, t) - {\textbf{j}}_m({\textbf{r}}, t) \right] \nonumber \\ &+& \varepsilon \int_{0}^{1}d\gamma p(\gamma)\tau^{\gamma-1}\frac{\partial^{\gamma} }{\partial t^{\gamma}}  \mathbf{E}({\textbf{r}},t)
\label{TF1}
\end{eqnarray}
is solenoidal,  as in the previous cases. If we consider $p(\gamma) = \delta(\gamma-1)$, then Eqs.~(\ref{Continuity-fractional1}) reduce to Eqs.(\ref{Continuity-normal}), as well as Eq.~(\ref{Jzero-order}) reduces to Eq.~(\ref{Jzero}). Other forms of $p(\gamma)$ may express a superposition of different diffusive regimes. For instance,
$p({\gamma}) = A \delta ({\gamma} -1) + B \delta({{\gamma}} - \alpha)$, where $A$ and $B$ are connected with  characteristic times, represents a superposition of a normal diffusion equation (${\gamma}=1$) with a diffusion equation of arbitrary order ${\gamma} = \alpha$, with $0 < \alpha \le 1$.

In the prototypical case of time-fractional derivative of order $\gamma$, as in Eq.~(\ref{Jzero-frac}), the displacement current assumes the form:
\begin{equation}
\label{jD}
{\textbf{j}}_D({\textbf{r}}, t) =  \varepsilon \tau^{\gamma-1}  \frac{\partial^{\gamma} \mathbf{E}({\textbf{r}},t) }{\partial t^{\gamma}}.
\end{equation}
For consistency, the definition of a modified displacement current
in terms of a fractional time-derivative operator requires an appropriated formulation of a time-fractional electrodynamics, which, needless to say, is a hard task. Recently, some approaches have been presented to deal with these kinds of formulation. Analogs for Maxwell's equations using time-fractional derivatives in the Riemann-Liouville and Caputo sense have been obtained by introducing time integro-differentiation of arbitrary order into the fundamental equations of the electrodynamics of a material media~\cite{Bogoliubov}.  The analysis shows that the stochastic nature of charged particles motion in a medium,
properly described in terms of fractional operators,  influences the dynamics of an electromagnetic field. More recently, a formulation of time-fractional electrodynamics was derived based on the Riemann-Silberstein vector~\cite{Stefanski2021}. The use of this vector together with  fractional-order derivatives indicates a way to write  Maxwell's equations in terms of time-fractional derivatives in a compact form, which allows for modeling of energy dissipation and dynamics of electromagnetic systems with memory. These generalizing approaches require a displacement current exactly in the form we are proposing here. They open the possibility to interpret the displacement currents, arising in the formulation of the time-fractional PNP model,  in the appropriate electrodynamics's framework.

The PNP model is a very useful framework to interpret impedance spectroscopy data regarding a wide variety of systems and electrolytic cells. In the range of frequency for which it works particularly well in describing the response of the ions in the sample to the stimulus of an external field, the electric current used to determine the electric impedance of the cell has to include the displacement current related to the time variation of the electric field. If the diffusion is normal, it is given by ${\bf j}_D=\varepsilon \partial {\bf E}/\partial t$. If the diffusion is anomalous it is given by (\ref{jD}). The important point is that, in the framework of anomalous diffusion the displacement current cannot be used in its usual, i.e., non-fractional form.  To analyze this aspect of the problem in more detail, let us consider a typical problem dealing with the impedance of an electrolytic cell. The proper application of the technique requires an electric field in the form:
\begin{equation}
\label{E-field}
{\textbf{E}}({\textbf{r}}, t) = {\textbf{F}}({\textbf{r}})e^{i \omega t},
\end{equation}
where $\omega = 2\pi f$ is the circular frequency of the applied field. From Eqs.~(\ref{Jzero}) and~(\ref{Continuity-frac}), and again using the equation of Poisson, Eq.~(\ref{Eq-Poisson}), we may write
\begin{eqnarray}
\nabla \cdot {\textbf{j}} =
 -\varepsilon \nabla \cdot \left[ \tau^{\gamma-1}
\frac{\partial^{\gamma}{\textbf{E}}({\textbf{r}},t)}{\partial t^{\gamma}}    -  \frac{\partial {\textbf{E}}({\textbf{r}}, t)}{\partial t}\right].
\end{eqnarray}
Now, if we use the definition of the electric field as in Eq.~(\ref{E-field}), we obtain:
\begin{equation}
\nabla \cdot {\textbf{j}} = -\varepsilon \left(\nabla \cdot {\textbf{F}}\right) (i \omega) \left[ (i \tau \omega)^{\gamma-1} - 1\right],
\end{equation}
which is solenoidal only if $\gamma=1$. However, if the difference
$ (i\tau \omega)^{\gamma-1} - 1$ may be considered negligible, the requirement of solenoidal current flowing through the system is essentially fulfilled, and the time-fractional approach to the PNP model may be fruitfully employed to analyze experimental data of impedance spectroscopy, as done in the last decades~\cite{Libro-Cambridge}. Some special attention has to be devoted to the $\omega \to 0$ limit, since $0\leq \gamma \leq 1$. Consequently the condition on the solenoidal character of ${\bf j}$ is violated in the dc limit.

In addition, when this condition is satisfied, a connection between the impedance spectroscopy response of an anomalous Poisson-Nernst-Planck (PNP) diffusional model and of equivalent circuits containing constant phase elements (CPEs) may be established in general terms for a typical electrolytic cell~\cite{JPCC-CPE}. The analysis, carried out in the limit of low frequency in order to highlight the surface effects, is shown to yield a natural presence of a CPE-like behavior in the diffusional model as analogous to what is considered as a tool used in the approaches based on equivalent electrical circuits.

\begin{figure}
\centering
 \includegraphics[width=0.500\textwidth]{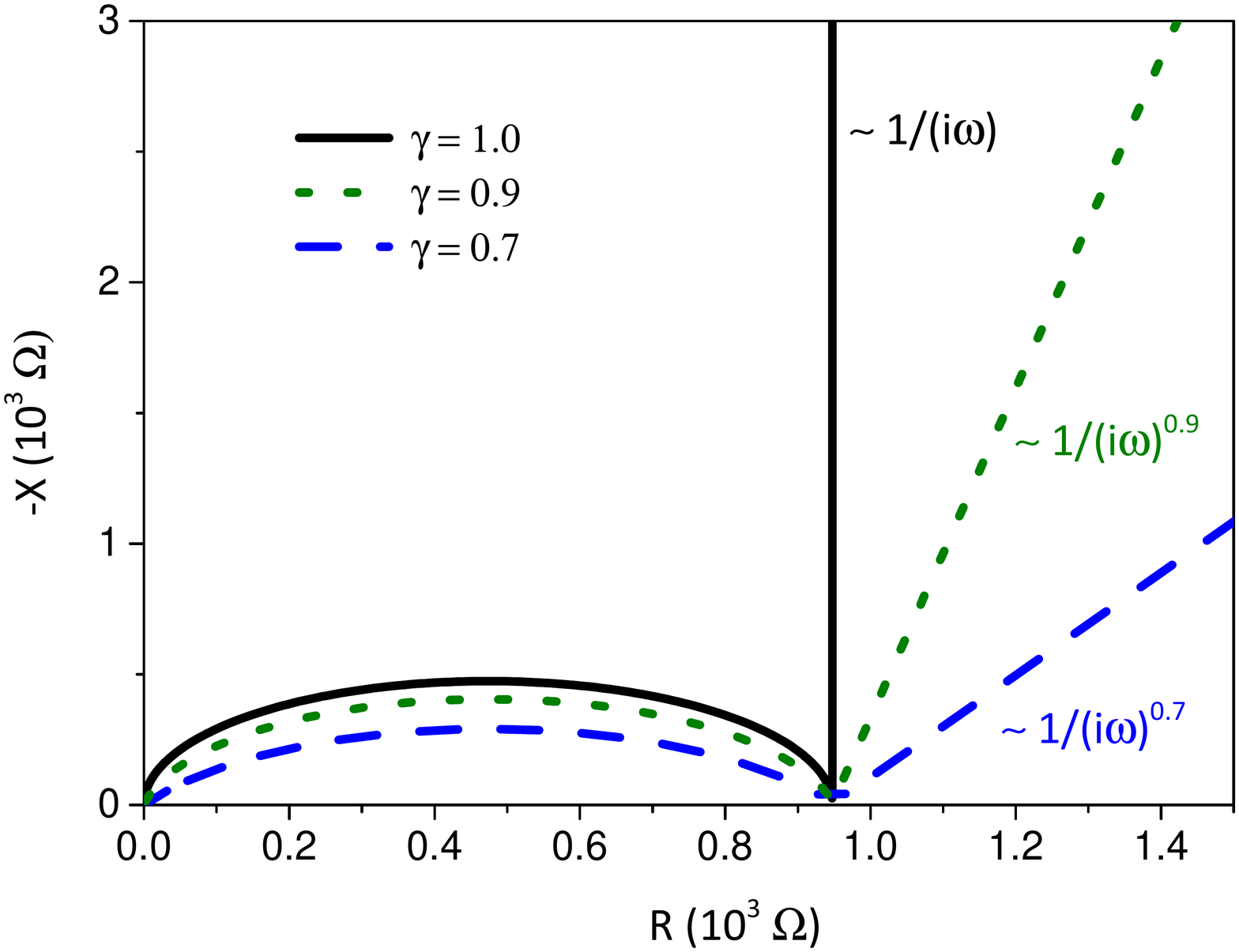}
\caption{Nyquist diagram for different values of $\gamma$,  with $\Phi(i\omega)=(i\omega)^{\gamma}$. The black (solid), green (dotted), and blue (dashed) lines correspond to the cases $\gamma=1.0$, $\gamma=0.9$, and $\gamma=0.7$, respectively.
We consider, for illustrative purposes, $D=8.0\times 10^{-8}\,$m$^2$/s, $A=3.1415\times 10^{-4}\,$m$^{2}$, $\varepsilon=90\varepsilon_{0}$ ($\varepsilon_{0}=8.85 \times 10^{-12}\,$ F/m),  $d=1.33 \times 10^{-3}\,$m, $\tau=1\, $s, and $\lambda=1.19\times 10^{-7}\, $m. }
\label{Fig1}
\end{figure}

\begin{figure}
\centering
\includegraphics[width=0.500\textwidth]{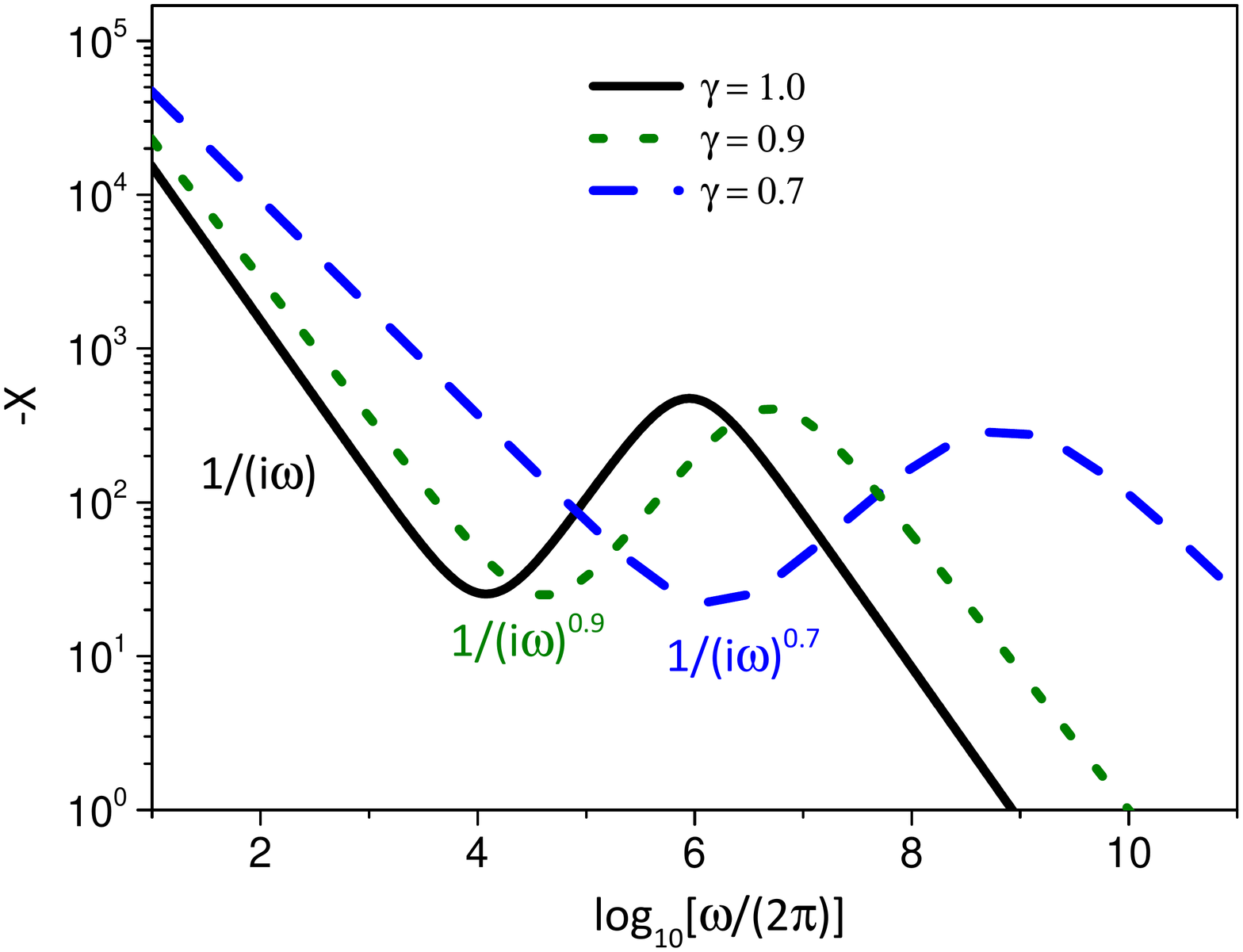}
\caption{The imaginary part of the impedance, $X = {\textrm{Im}}(Z)$,  for different values of $\gamma$, with $\Phi(i\omega)=(i\omega)^{\gamma}$. The black (solid), green (dotted), and blue (dashed) lines correspond to the cases $\gamma=1.0$, $\gamma=0.9$, and $\gamma=0.7$, respectively.
The parameters are the same as in Fig.~\ref{Fig1}. }
 \label{Fig2}
  \end{figure}

\begin{figure}
\centering
\includegraphics[width=0.500\textwidth]{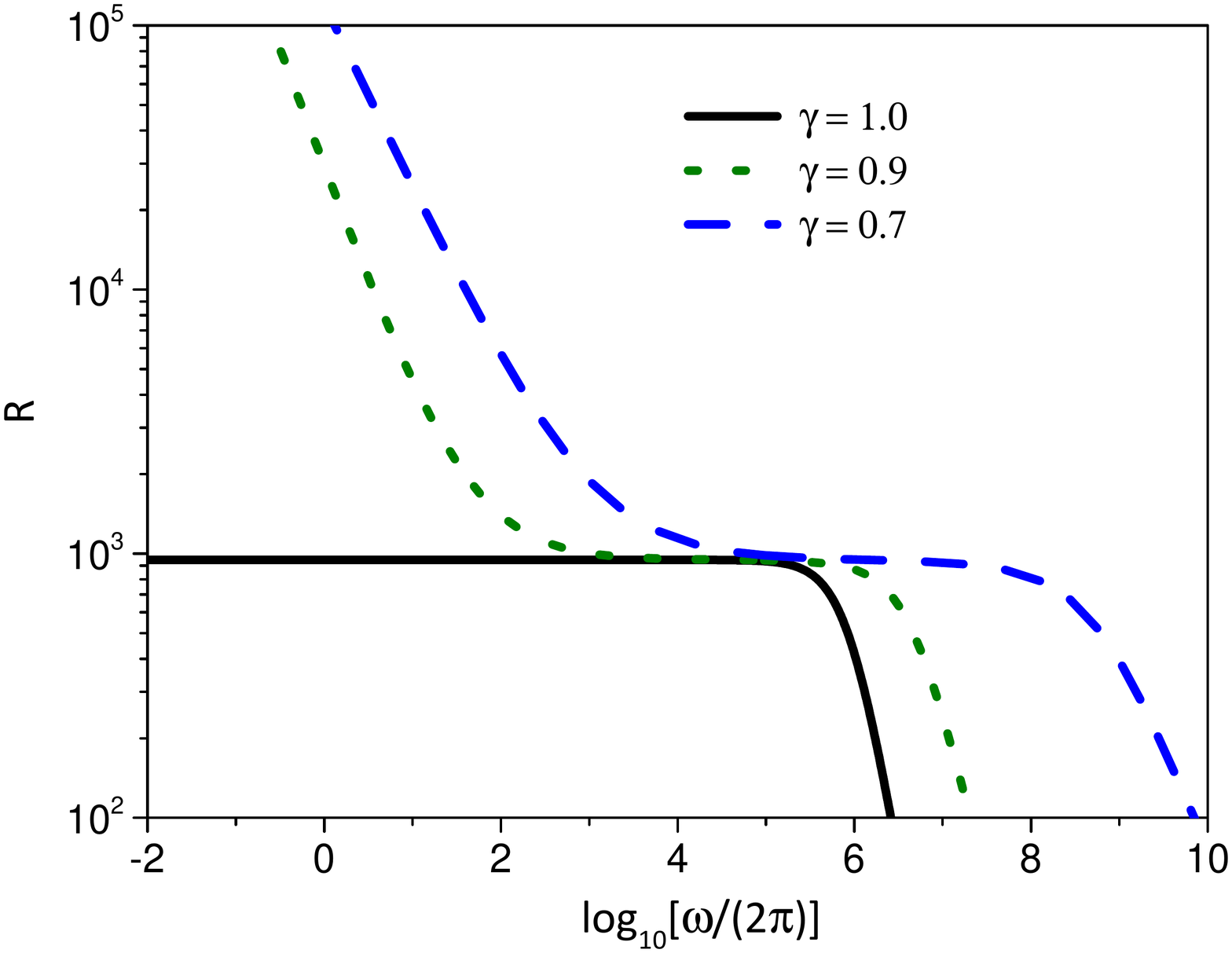}
\caption{The real part of the impedance, $R = {\textrm{Re}}(Z)$,  for different values of $\gamma$, with $\Phi(i\omega)=(i\omega)^{\gamma}$. The black (solid), green (dotted), and blue (dashed) lines correspond to the cases $\gamma=1.0$, $\gamma=0.9$, and $\gamma=0.7$, respectively.
The parameters are the same as in Fig.~\ref{Fig1}. }
 \label{Fig3}
  \end{figure}

A simple, but representative, example is in order here to illustrate this important tool provided by the extensions of the PNP model involving fractional operators. To explore it, we first define the electrolytic cell, i.e., the region where the charges are diffusing,  in the shap of a slab o thickness $d$ in a Cartesian frame with the electrodes positioned at $z =\pm d/2$, where $z$ the coordinate normal to the surfaces,  which have area $A$. This way, the problem becomes one-dimensional. For simplicity, we consider also the case of full dissociation, and we assume that the ions have the same diffusion coefficients. In this way, the equations to be solved are obtained from Eqs.~(\ref{Continuity-fractional1}), rewritten as
\begin{eqnarray}
\label{Continuity-fracc1}
\int_{0}^{1}d\gamma p(\gamma)\tau^{\gamma-1}\frac{\partial^{\gamma}}{\partial t^{\gamma}} n_p(z, t) +\frac{\partial}{\partial z} j_p(z, t) &=&0,  \nonumber\\
\int_{0}^{1}d\gamma p(\gamma)\tau^{\gamma-1}\frac{\partial^{\gamma}}{\partial t^{\gamma}}n_m(z, t) + \frac{\partial}{\partial z} j_m(z, t) &=& 0,
\end{eqnarray}
together with the equation of Poisson, Eq.~(\ref{Eq-Poisson}).
The cell is subjected to a  time-dependent  electric potential $V\left(\pm d/2,t\right)=\pm \left(V_{0}/2\right) e^{i\omega t}$ and we limit our considerations to the {\bf periodic}-state.  For perfect blocking electrodes, i.e., when the boundary conditions $j_{p}(\pm d/2,t)=j_{m}(\pm d/2,t)=0$ hold, the electrical impedance spectroscopy may be investigated in the asymptotic limit of small AC-signal limit. This condition permits one to search for solutions of the fundamental equations, Eqs.~(\ref{Continuity-fracc1}) and (\ref{Eq-Poisson}) in the form
\begin{equation}
n_{p}(z,t)=N+\eta_{p}(z)e^{i\omega t} \quad {\textrm{and}}\quad n_{m}(z,t)=N+\eta_{m}(z)e^{i\omega t}
\end{equation}
with $N>> |\eta_{p}(z)|$ and $N>> |\eta_{m}(z)|$, where
$N$ represents the number of ions per unit  volume in equilibrium,
and $V(z,t)=\phi(z)e^{i\omega t}$ (see Ref.~\cite{Libro-Cambridge}, for details). We  obtain for the electrical impedance the following analytical expression~\cite{JPCC-CPE}:
\begin{eqnarray}
\label{Impedance1}
Z=\frac{2}{\Phi(i\omega) A \varepsilon \beta^{2}}\left\{
\frac{1}{\lambda^{2}\beta}\tanh\left(\beta  d/2\right)+\frac{d}{2{D}}\Phi(i\omega) \right\},
\end{eqnarray}
where $\beta=\left(1/\lambda\right)\sqrt{1+\Phi(i\omega)\lambda^{2}/D}$, $\Phi(i\omega)=(1/\tau)\int_{0}^{1}d\gamma p(\gamma)(i\omega\tau)^{\gamma}$, and
$\lambda=\sqrt{\varepsilon k_{B}T/\left(2q^{2}N\right)}$ is the Debye screening legnth.
Figures~\ref{Fig1},~\ref{Fig2}, and~\ref{Fig3} show some prediction obtained from Eq.~(\ref{Impedance1}) for different values of $\gamma$. We observe that in the low frequency limit the fractional time derivative plays an important role and is responsible for a different behavior from the normal one. It is characterized by $Z\sim 1/(i\omega)^{\gamma}$, which has a direct connection with the presence of constant phase elements (CPE) when the problem is formulated in terms of equivalent circuits~\cite{JPCC-CPE}.

This can be checked by considering in more detail the asymptotic behavior of the impedance in the low frequency limit, which,  for Eq.~(\ref{Impedance1}),  is given by
\begin{equation}
\label{23}
Z\approx \frac{\lambda^2 d}{\varepsilon A D}+\frac{2\lambda}{\varepsilon A}\frac{1}{\Phi(i\omega)}\;.
\end{equation}
The first addendum in Eq.~(\ref{23}) is directly related to the resistance of the cell, it is a bulk contribution and it scales with the thickness of the cell, $d$. The other term depends on frequency, contributing to the real and imaginary parts of the impedance in the low frequency limit similar to the CPE contribution to the total impedance of the cell. It takes origin from the surface layer whose thickness is of the order of $\lambda$. In the case of normal diffusion, it reduces to a surface capacitance. If, instead,  the diffusion is anomalous, the corresponding term takes into account also dissipative effects, introducing a surface electrical resistance.  In particular, for the cases illustrated in  Figs.~\ref{Fig1},~\ref{Fig2}, and~\ref{Fig3},  where $\Phi(i\omega)=(i\omega)^{\gamma}$,  the contribution to the impedance of the cell is $1/(i \omega)^{\gamma}=\omega^{-\gamma}[\cos(\gamma \pi/2)-i \sin(\gamma \pi/2)]$ as predicted by the constant phase element~\cite{Pajkossy,Liu,Jorcin}.
The suitable consideration for the displacement current lead us to a different behavior from the one obtained by considering the standard form of the displacement current~\cite{JPCB113}, which implies in the following asymptotic limit
\begin{eqnarray}
\label{24}
Z\approx \frac{\lambda^2 d}{i\omega\varepsilon A D}\Phi(i\omega)+\frac{2\lambda}{i\omega\varepsilon A}\;,
\end{eqnarray}
with a different behavior for the impedance. In particular, the behavior related to the constant phase element present in Eq.~(\ref{23}) is absent in Eq.~(\ref{24}) for the imaginary part of the impedance in the very low frequency limit.

To sum up, we have analyzed the role of the displacement current in a generalized time-fractional approach to the PNP model for electrical impedance. We have shown that, in order to keep the concept of electrical impedance meaningful, the total electric current arising in this extended model has to be solenoidal. This requirement implies that also the expression of the displacement current has to be modified by extending it to the fractional domain too.  The proposed form of the displacement current coincides with the one used recently in some generalizations of the Maxwell equations for material media involving fractional order operators. The time-fractional extension of PNP model naturally exhibits a CPE-like behavior of the electrical impedance in the low frequency region of the spectra and is shown to be a powerful tool to face a wide variety of complex behaviour found in electrolytic cells. From the analysis reported above, we conclude that the CPE behaviour observed in the frequency dependence of the impedance of electrolytic systems, can have different origin~\cite{Jorcin}. However, in the case of media presenting a porous structure, it could be related to the anomalous diffusion of the ions in it.

\textbf{Acknowledgements.} This work was supported by the MEPhI Academic
Excellence Project (G. B.)
and by the Program of Visiting Professor of Politecnico di Torino (L. R. E.). E.K.L. thanks partial financial support of the CNPq under Grant No. $302983/2018-0$.


\end{document}